\documentclass[a4paper]{article}
\title{Guaranteed successful strategies for a square
achievement game on an n x n grid}
\author{Thomas Jenrich}
\date{2012-05-20}
\begin{document}
\maketitle

\section{Abstract and introduction}

At some places (see the references) Martin Erickson describes a
certain game:

``Two players alternately write O's (first player) and X's (second
 player) in the unoccupied cells of an $n$ x $n$ grid. The first player (if
 any) to occupy four cells at the vertices of a square with horizontal
 and vertical sides is the winner.''

Then he asks

``What is the outcome of the game given optimal play?'' or

``What is the smallest $n$ such that the first player has a winning
 strategy?''

For $n$ lower than 3 a win is obviously impossible.

The aim of this article and the additionally (in the source package)
provided computer program SQRGAME2 (as a revision of its incorrect
predecessor SQRGAME) is to give and prove sure strategies for the
second player not to lose if $n$ is 3 or 4, and for the first player
to win if $n$ is 5.

This article has been updated only because it turned out that SQRGAME was
incorrect. You can find the details in the paragraph beginning with
``The predecessor SQRGAME ''
within the section ``The provided computer program SQRGAME2''.

\section{Reformulation and analysis of the problem}

To prepare the use of a computer program to check all possible cases
this description of the game is given:

Let $G$ be a two-dimensional array variable with integer index ranges
from 0 to (at least) $n-1$ and possible cell values 0, 1 and 2.

The current value of a cell addressed by a row number $r$ and a column
number $c$ is denoted by $G[r,c]$ where $r$ and $c$ have to be integer values
from 0 to $n-1$, sometimes presented as a position ($r$,$c$).

The game starts with $G[r,c] = 0$ for all integers $r$ and $c$ from 0 to $n-1$.
Player 1 and player 2 alternately change the value of a single cell of
$G$ at some permitted position from 0 to 1 or 2, respectively.

For $v=1$ or $v=2$, player $v$ wins the game if he is the first to achieve
that there are non-negative integers $r$ and $c$ and a positive integer $d$
such that

\hspace{0.1in}$r+d<n$, $c+d<n$, $G[r,c]=v$, $G[r,c+d]=v$, $G[r+d,c]=v$, and $G[r+d,c+d]=v$.

That winning four-cells-configuration will be shortly called a square
(of size $d$) here and in the source code of the provided computer
program.

It is easy to verify that rotating the content of G by steps of 90
degrees or mirroring it with respect to the columns, the rows, or one of
the diagonals preserves existing squares (and their sizes) and does not
establish new ones.
\vspace{0.1in}

Because we want to consider only optimal playing and try to reduce the
lengths and number of different game runs we require:

When a player has to make his move:

\hspace{0.1in} If he can win by completing a square he wins instantly.

\hspace{0.1in} Otherwise:

\hspace{0.2in}  If there are two (or more) different positions his opponent could use

\hspace{0.2in}  with his next move to complete a square (a situation further called a

\hspace{0.2in}  dilemma) the currently active player loses instantly.

\hspace{0.2in}  Otherwise:

\hspace{0.3in}    If there is one such position the currently active player has to

\hspace{0.3in}    occupy that position.

\hspace{0.3in}    Otherwise:

\hspace{0.4in}     If the currently active player can complete a dilemma (against his

\hspace{0.4in}     opponent) he wins instantly.

\hspace{0.4in}     Otherwise:

\hspace{0.5in}      As long as the content of $G$ is symmetric with respect to the

\hspace{0.5in}      columns (meaning $G[r,c]=G[r,n-1-c]$ for all allowed positions ($r$,$c$))

\hspace{0.5in}      no move is allowed using a column number $c>((n-1)$ $div$ 2), where $div$

\hspace{0.5in}      is the infix operator delivering the integer part of the quotient

\hspace{0.5in}      of the two operands.

\hspace{0.5in}      $G$ is obviously symmetric at the start of the game and stays

\hspace{0.5in}      symmetric as long as there are only moves using the column number

\hspace{0.5in}      ($n-1)/2$ , requiring $n$ to be odd.

\hspace{0.5in}      For the position ($r$,$c$) of the first move of the game, $r$ is not

\hspace{0.5in}      allowed to be greater than $c$. If $n$ is odd and the first move has

\hspace{0.5in}      used the middle cell at position ($n$ $div$ 2, $n$ $div$ 2) that applies

\hspace{0.5in}      also to the second move.
\vspace{0.1in}

For $v=1$ or $v=2$, player $v$ can not win if no possible square having no
vertex occupied by his opponent is left. So in this case one can stop
the game instantly if it is the aim to show whether player $v$ can win or
his opponent can prevent a loss.

\section{The provided computer program SQRGAME2}

For some given $n$ (in the current version 3, 4, and 5) SQRGAME2 is used
to check whether player 1 can win or player 2 can prevent to lose. It
performs a search using a backtracking algorithm to investigate all
possible game runs observing the restrictions given above. It does not
try to find out whether player 2 could win.

The program integrates additional restrictions to player 2 in the cases
$n=3$ and $n=4$ and additional restrictions to player 1 in the case $n=5$.

The execution of the program will show that player 2 reaches his aim in
the cases $n=3$ and $n=4$, and player 1 reaches his aim in the case $n=5$,
while the successful players don't perform any actual backtrackings.

In the case $n$=4 player 2 and in the case $n$=5 player 1 is required to
occupy a free position that now or later could become a vertex of a
possible square for player 1 if such a position exists and the prior
position selection rules don't apply.

In the case $n=5$ the maximum number of moves in a game is restricted:

Instead of the natural limit, the number of cells $n^2 = 25$, much smaller
values are used: If the second move uses the position (0,2) the maximum value
is set to 17, if the position is (0,0) or (1,2) to 13, otherwise to 11.

These limitations greatly reduce the number of game runs to be checked
when the explicit position selection given in the routine \emph{p1\_n5\_fix\_pos}
does handle only the case \emph{n\_moves}=0.

Actually the positions given in \emph{p1\_n5\_fix\_pos} for the even values from
2 to 12 for \emph{n\_moves} are taken from observations of game runs with
explicitly handling only lower values for \emph{n\_moves}, therefore causing
actual backtrackings.

The explicit selections and restrictions mentioned above combined with the
default search orders given in the routines \emph{player\_1\_tries\_to\_win} and
\emph{player\_2\_tries\_not\_to\_lose} \\ establish constructive strategies.
\vspace{0.1in}

The provided source code has a fairly simple structure. The included
comments (enclosed in curly braces) should be sufficient at least for
readers knowing at least one imperative programming language.

Because case statements are used a lot I should mention this:

Pascal case statements are very similar to switch statements in C but:

Instead of the keyword \emph{switch} the keyword \emph{case} is used. The keyword
\emph{case} to start a case is omitted and there is an implicit \emph{break} at the
end of the statement part of each case. (Those are additional reasons
not to convert the program into C.)
\vspace{0.1in}

The predecessor SQRGAME had to be revised because the contained line

\vspace{0.05in}
\ttfamily
 or (n=5) and can\_complete\_a\_dilemma(1,r,c)
\rmfamily
\vspace{0.05in}

within the routine \emph{player\_2\_tries\_not\_to\_lose} required player 2
in the case $n=5$ to occupy a cell in order to disable player 1 to occupy
that cell and completing a dilemma (against player 2) with his next move.

But sometimes player 2 can reach a better result (a draw instead of a loss)
if he occupies another cell and that way form an almost complete square,
causing player 1 to use his next move to prevent that square (instead of
completing the mentioned dilemma).

After that correction SQRGAME computed a draw (for player 1) in the case $n=5$.
But I was able to change the routine \emph{p1\_n5\_fix\_pos} and the settings of
\emph{max\_n\_moves} (the maximum limit of the number of moves in a game) such that
SQRGAME2 again computes a win for player 1 in the case $n=5$.

\vspace{0.1in}

To compile the provided source code file SQRGAME2.PAS you will need a
compiler compatible with Turbo Pascal 4.0.

I have tested it with Turbo Pascal 5.5 and 7.01 under MS-DOS 5.0 and
MS-Windows 98SE, with Borland Delphi 4.0 build 5.37 under MS-Windows
98SE, and with Virtual Pascal 2.1 build 279 under MS-Windows 98SE.

The use of Free Pascal should also be possible but I haven't tested that.

To avoid a compilation result depending on the settings you could use
the command line versions of the compilers (TPC for Turbo Pascal, BPC
for Borland Pascal 7, DCC32 for Borland Delphi (32 bit versions; do not
miss to use the -CC option in order to generate a console executable),
VPC for Virtual Pascal) instead of the compilers integrated in the IDEs.

\vspace{0.1in}
The program does not use the heap or any pointer operation at all.
If you don't change the respective compiler directives, range checks
and stack overflow checks are generated. So the resulting executable
will be extremely safe. It is also very small and needs only a few
kilobytes for the stack.

For the current version running on a 1 GHz Intel PIII under Windows 98SE
takes less than one second if Borland Delphi 4.0 or Virtual Pascal 2.1
is used, less than three seconds if Turbo Pascal 5.5 or 7.01 is used.

Running on a 500 MHz AMD K6-II under MS-DOS 5.0 takes about four seconds
if Turbo Pascal 5.5 or 7.01 is used.

The program ignores any command line parameters or inputs other than
pressing Ctrl-C to cancel the execution.

It writes only to the standard output device. In the default case that
will be the monitor screen. But you can redirect the output to a file.
That way the lines below enclosed in $<<<$ and $>>>$ were generated. You
could use them to compare your results with.
\vspace{0.1in}

$<<<$
\ttfamily
\small

=== Checking solutions for the square achievement game problem ===

===     Version 2      Copyright (c) 2012-05-20 Thomas Jenrich ===
\vspace{0.1in}

Hints:

\hspace{0.05in} After each 100000 moves a + will be emitted.

\hspace{0.05in} To cancel the execution press Ctrl-C .
\vspace{0.2in}

Starting search with n = 3
\vspace{0.1in}

The search has been completed. Result: draw

Sum of moves: 76.  N. of backtrack.: Player 1: 30, Player 2: 0
\vspace{0.2in}

Starting search with n = 4

+

The search has been completed. Result: draw

Sum of moves: 112751.  N. of backtrack.: Player 1: 47595, Player 2: 0
\vspace{0.2in}

Starting search with n = 5
\vspace{0.1in}

The search has been completed. Result: win

Sum of moves: 7974.  N. of backtrack.: Player 1: 0, Player 2: 4713
\vspace{0.1in}

== Regular program stop ==

\rmfamily

$>>>$

\section{References}

Martin Erickson, \emph{Open Problems}

http://www2.truman.edu/\verb+~+erickson/openproblems.html
\vspace{0.1in}

Martin Erickson, \emph{Square achievement game on an n x n grid}

http://garden.irmacs.sfu.ca/?q=op/a\_game\_on\_an\_n\_x\_n\_grid
\vspace{0.1in}

Martin Erickson, \emph{Square Achievement Game on a Grid}

http://mathoverflow.net/questions/29608/square-achievement-game-on-a-grid

\vspace{0.1in}

\emph{Turbo Pascal} versions 1.0, 3.02, and 5.5 (binaries only)

http://edn.embarcadero.com/museum/antiquesoftware\#

For downloading one has to register or sign-in.

\vspace{0.1in}

\emph{Virtual Pascal} (Closed Source freeware)

One ZIP-file including binaries and documentation for Win32, OS/2, and Linux

Official forum:

http://vpascal.ning.com/

Forum entry \emph{Where can I download VP?} :

http://vpascal.ning.com/forum/topic/show?id=854411\verb+%3ATopic%3A9+

\vspace{0.1in}

\emph{Free Pascal} (Open Source freeware)

Sources, documentation, and binaries for several systems

http://www.freepascal.org

\end{document}